# The Materials Simulation Toolkit for Machine Learning (MAST-ML): an automated open source toolkit to accelerate data-driven materials research

Ryan Jacobs[1], Tam Mayeshiba[1], Ben Afflerbach[1], Luke Miles[2], Max Williams[3], Matthew Turner[2], Raphael Finkel[2], Dane Morgan[1]

[1] Department of Materials Science and Engineering, University of Wisconsin-Madison, Madison, WI, 53706, USA

[2] Department of Computer Science, University of Kentucky, Lexington, KY, 40506, USA

[3] Computer Engineering and Computer Science Department, University of Louisville, Louisville, KY 40292, USA

**Abstract:**

As data science and machine learning methods are taking on an increasingly important role in the materials research community, there is a need for the development of machine learning software tools that are easy to use (even for nonexperts with no programming ability), provide flexible access to the most important algorithms, and codify best practices of machine learning model development and evaluation. Here, we introduce the Materials Simulation Toolkit for Machine Learning (MAST-ML), an open source Python-based software package designed to broaden and accelerate the use of machine learning in materials science research. MAST-ML provides predefined routines for many input setup, model fitting, and post-analysis tasks, as well as a simple structure for executing a multi-step machine learning model workflow. In this paper, we describe how MAST-ML is used to streamline and accelerate the execution of machine learning problems. We walk through how to acquire and run MAST-ML, demonstrate how to execute different components of a supervised machine learning workflow via a customized input file, and showcase a number of features and analyses conducted automatically during a MAST-ML run. Further, we demonstrate the utility of MAST-ML by showcasing examples of recent materials informatics studies which used MAST-ML to formulate and evaluate various machine learning models for an array of materials applications. Finally, we lay out a vision of how MAST-ML, together with complementary software packages and emerging cyberinfrastructure, can advance the rapidly growing field of materials informatics, with a focus on producing machine learning models easily, reproducibly, and in a manner that facilitates model evolution and improvement in the future.

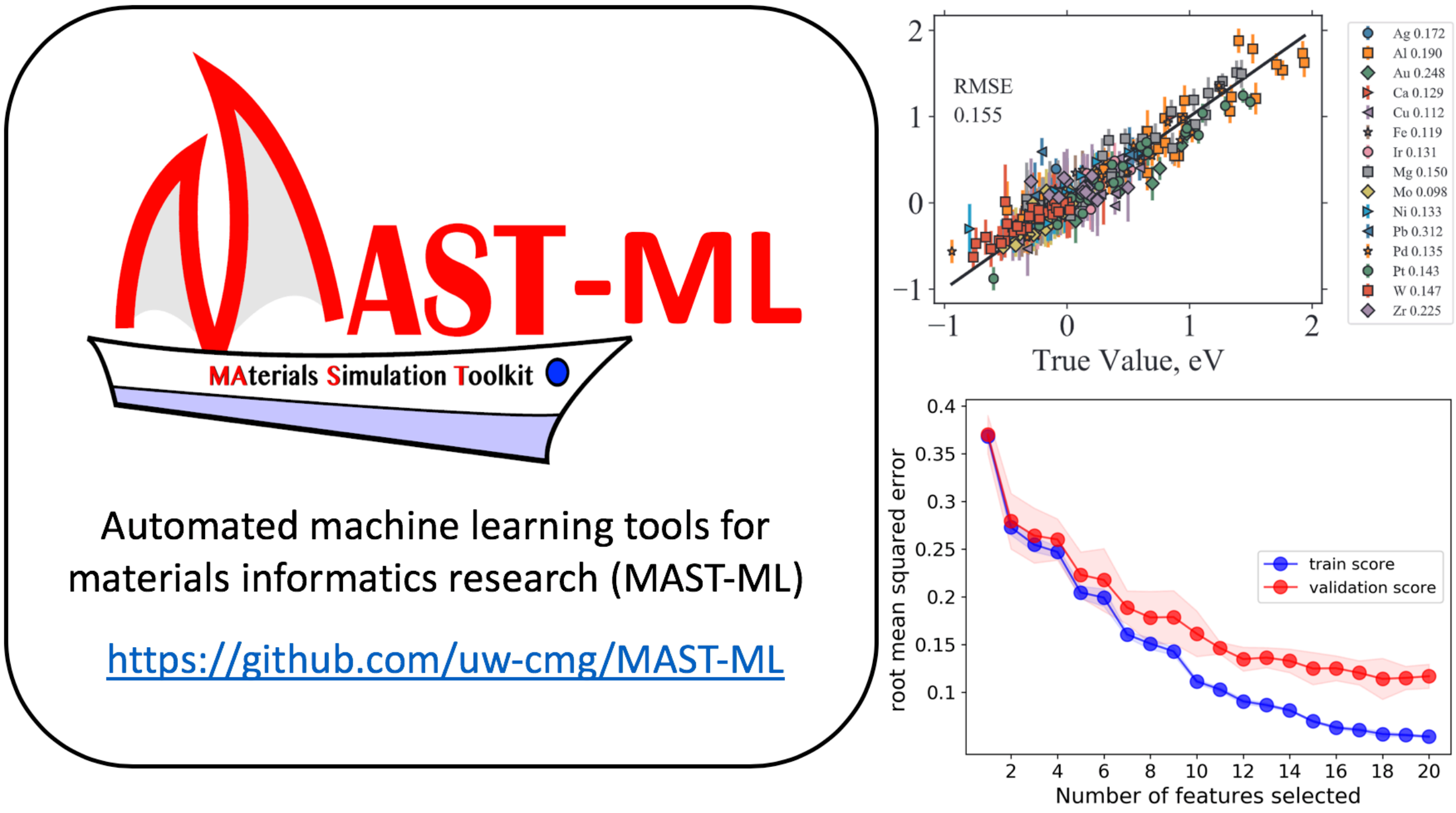

## 1. Introduction:

The Materials Genome Initiative[1,2] was instrumental in creating troves of available materials data contained in databases such as the Materials Project, the Open Quantum Materials Database and Citrination.[3–8] This unprecedented scale of available materials data, coupled with advances in computing hardware (e.g. development of GPU clusters), development of novel algorithms (e.g. deep convolutional neural networks) and streamlined, open-source availability of software streamlining the use of machine learning algorithms (e.g. the scikit-learn package) have enabled the combined data science and machine learning (often called materials informatics) subfield of materials research to explode over the past five years, with a number of overviews and reviews having been written.[9–16] The recent application of machine learning in materials science has been broad, including, e.g., unsupervised learning analysis of X-ray diffraction and electron microscopy images,[17–19] deep learning applied to microstructural image recognition of defects and predictions of material stability,[20–22] high-throughput autonomous experiments,[23–25] text mining with natural language processing to inform materials synthesis and processing procedures,[26,27] and active learning techniques applied to improve machine learning potentials and targeted materials design.[28,29] Due to their ubiquity and relevance in establishing predictive processing-structure-property-performance relationships relevant for materials discovery and design, in this paper and the development of MAST-ML to date we place a particular emphasis on the use of supervised machine learning models for the purpose of regression tasks. Supervised machine learning regression models have been formulated for many classes of materials spanning an array of materials properties and associated applications, with some recent studies including the prediction of dielectric properties of perovskites and polymers,[30,31] bulk stability of perovskite and garnet materials,[20,32] electronic bandgap of numerous types of inorganic materials,[33–36] superconducting critical temperatures,[37,38] electromigration in metals,[39] dilute solute diffusion barriers in metals,[40,41] scintillator material discovery,[42] and melting points of unary and binary solids,[43] among many others.

Supervised machine learning regression seeks to solve the general problem of finding the function $F$ in $Y = F(X)$ from a known feature matrix $X$ and target values $Y$. $F$ is found by machine learning typically such that the lowest errors between the predicted and true $Y$ values are realized. While many machine learning algorithms are readily available in open-source packages such as scikit-learn[44] (sometimes called sklearn) and Keras[45] (based on TensorFlow[46]) to investigate such supervised learning problems, the overall supervised machine learning workflow for materials problems is still largely executed by hand, meaning individual users and research groups tend to develop and use their own in-house methods and scripts to manage all the steps outside those of the core algorithm. This practice introduces significant barriers to entry for new researchers, particularly non-experts in machine learning, which barriers are typically the result of the numerous technical and practical intricacies of successfully formulating and evaluating machine learning models in materials informatics research problems. In addition, this practice may also make model reproducibility and model evolution (e.g. improvement in predictions given additional training data in the future) difficult. In this paper, we introduce the Materials Simulation Toolkit for Machine Learning (MAST-ML).[47] MAST-ML is an open source Python-based software package designed to broaden and accelerate the use of machine learning in materials science research. MAST-ML enables users to rapidly develop machine learning models, codifies best practices in standard supervised learning workflows, and seeks to lower the barrier for non-experts to perform materials informatics research by relying on minimal to no programmatic input from the user.

As a concrete example of how MAST-ML could impact the materials informatics community, consider the task of building and evaluating a machine learning model for predicting values of electronic bandgaps and the time spent by a typical researcher relatively new to the materials informatics field to formulate and evaluate such a model. With the tools and methods available today, one might spend weeks searching papers, discovering digital repositories, and learning to use relevant Application Programming Interfaces (APIs) to pull in data, and then weeks developing features, perhaps pulling from databases of elemental properties and/or structural databases and finding or developing tools to extract fingerprints of structures appropriate for the machine learning problem at hand. The machine learning model development might then take a few months as the researcher learns and applies best practices to data pre-processing, explores multiple machine learning model algorithms, and assesses the results. Much of the time would be spent on relatively minor but time-consuming tasks like typing in missing elemental property features, coding different ways of splitting the data for cross validation, hyperparameter optimization, and discovering approaches and best practices of which they might not be aware (e.g., stratified vs. standard k-fold cross validation). The final model would be used to predict bandgaps for some set of compounds of interest, e.g., new III-V semiconductors, perhaps being used to support a further exercise in materials development, and then be published as a standalone paper. On the other hand, this same materials informatics research project, which might take a relatively new researcher many months to execute, could easily be done much faster using the tools contained in MAST-ML. While any materials informatics project typically involves multiple iterations of analysis and re-evaluation of, e.g., the feature set used or the model employed, MAST-ML allows researchers to quickly and easily conduct this materials informatics research loop, thus dramatically lowering the technical barrier and time required to produce meaningful results and analysis using machine learning for materials research problems.

Recently, there has been intense development of open source software packages (in addition to MAST-ML) aimed at streamlining and accelerating the adoption of materials informatics research. These packages include, e.g., Lolo (as implemented in Citrination),[8,48]

AFLOW-ML,[49] matminer,[50] the Materials Knowledge Systems in Python project (pyMKS),[51] veidt,[52] and the Materials Agnostic Platform for Informatics and Exploration (MAGPIE),[53] among others. MAST-ML fits into the broader ecosystem of these materials informatics tools by either being complementary, symbiotic, or supportive of each of the above software packages. For example, MAST-ML currently integrates key functionality of both MAGPIE and matminer to enable users to create an extensive initial feature matrix to evaluate machine learning models and easily pull in tabulated materials data from online databases such as the Materials Project.

In this paper, we discuss (i) the core components of the supervised machine learning workflow and key capabilities of MAST-ML in **Section 2**, (ii) how to obtain and run MAST-ML in **Section 3**, (iii) the key data and input files for a MAST-ML run in **Section 4** and **Section 5**, respectively, and (iv) the MAST-ML run output structure and key contents in **Section 6**. **Section 7** demonstrates some recent examples of the type of research enabled by this software package. Finally, **Section 8** provides a roadmap for how we envision this and related software may evolve in the greater materials informatics ecosystem as the use of data-driven techniques become increasingly prevalent in materials research.

## 2. The MAST-ML workflow:

The focus of the development of MAST-ML was to automate the process of conducting supervised machine learning problems, with a particular emphasis on regression problems in materials science research. The overview of such a supervised machine learning problem is presented in **Figure 1**. In **Figure 1**, each component of the supervised learning workflow is shown in a particular block, and the text within each block denotes examples of operations (not an exhaustive list) one may conduct with MAST-ML as part of that portion of the workflow. Each component of the supervised learning workflow and MAST-ML-specific details for each component are enumerated below. The roman numerals labeling each component in this list are also used to label the respective workflow components in **Figure 1**. In addition, these roman numeral labels coincide with the MAST-ML input file sections discussed in **Section 5** and the output data file structure discussed in **Section 6**.

The user may specify multiple types of a given operation at each step (for example, multiple feature normalization methods, feature selection techniques, machine learning models, and cross validation data splitters) in a single input file, and thus execute many types of model development and assessment in a single MAST-ML run. Based on the supervised learning workflow illustrated in **Figure 1** and discussed in this section, MAST-ML iterates through every combination of specified feature normalization, selection, machine learning model, and cross validation data split within a single run. Therefore, if the user specifies two methods of feature normalization, two methods to select features, four machine learning models, and three methods of cross validation, this specification amounts to a total of 2 × 2 × 4 × 3 = 48 different supervised learning workflows to be executed within the run.

(I) *Data import.* A data file (either in .csv or .xlsx format) is supplied by the user when a MAST-ML run is started. Information is provided for which column(s) of the data file denote values of the feature matrix $X$ and target data $Y$ on which to train a machine learning model. In addition, the user may specify other columns of the data file to denote one or more of the following: (i) grouping - information used in data subsampling (e.g. group labels for leave out group cross-validation), (ii) test - data restricted to only function as test data (i.e. data that is never used in model training or validation), (iii) comments - additional comment information useful for one to include in the data file for organizational purposes but is not to be used in model training in any way (e.g. general comments on specific data points). Additional information regarding the data file can be found in **Section 4**.

(II) *Data cleaning.* Data used in machine learning problems is often imperfect and may contain missing, unimportant, or incorrect values. Missing data: points missing some feature values can be removed or data imputation methods can be used to fill in approximate values for the missing data entries. Currently, methods to treat missing data include removal of missing values, imputation by mean, median and mode value, and approximation of the missing value using principal component analysis (PCA). Note that the use of PCA in this context also constitutes a form of data imputation, however we have separated it for practical use reasons because the imputation methods used by MAST-ML are called directly from the scikit-learn package. Unimportant data: Currently a feature is identified as unimportant if it has a constant value for all rows in the data file and can be automatically removed as such features would be unlikely to add predictive value to a model. Incorrect data: Currently we have provided a feature which automatically examines the input data values for each *X* and *Y* data column and flags input values which may be problematic based on their values being more than two standard deviations from the average value of that particular data column. These potentially problematic values are not removed but are noted in a dedicated output file.

(III) *Feature matrix generation.* In addition to feature matrix values specified in the imported data file supplied by the user, one can also use certain prescribed methods to generate additional features. If provided information of the material composition (given as a labeled column in the input data file) for each target data point, MAST-ML can construct a large set of element-based feature properties following the MAGPIE approach.[54] The user can also automatically query materials databases such as the Materials Project[4] with provided material compositions to search for relevant data for those specific compositions, e.g., relative stability to other compounds. Finally, if information on the material composition and structure is provided in the input data file (specifically, in the form of a pymatgen[55] structure object), a suite of different structural features can also be generated. These structural features are sometimes referred to as structural fingerprints, and consist of features generated using techniques such as the smooth overlap of atomic orbitals,[56] the coulomb matrix,[57] bag of bonds,[58] etc. These structural features are generated using the matminer package.[50]

(IV) *Feature matrix normalization.* It is common practice to scale the values of the feature matrix in a manner such that features with disparate ranges (e.g. one feature ranging from 1 to 10 and another from -1000 to 1000) do not unphysically impact model training. Here, we mainly draw upon the feature normalization routines in scikit-learn. In addition, we have provided a custom feature normalization routine for MAST-ML, which allows the user to specify a particular mean and standard deviation value for which to scale the values of the feature matrix.

(V) *Feature selection.* To formulate an accurate machine learning model, one typically wants to achieve the highest model accuracy using the least number of features possible, thus resulting in the simplest model. This is typically done by selecting a subset of the feature matrix that minimizes some scoring function, e.g. the root mean square error (RMSE) from 5-fold cross validation, to simultaneously minimize the model error and minimize possible overfitting. MAST-ML leverages the feature selection routines contained in scikit-learn as well as the forward selection routine contained in the mlxtend package.[59] In addition, a custom forward selection routine was written for MAST-ML, which offers additional user flexibility compared to routines contained in standard packages regarding the model and cross validation method used to select features.

(VI) *Learning curve construction.* Learning curves show details of the feature selection process as the number of features and amount of training data changes. MAST-ML enables the construction of two types of learning curves. The first type of learning curve plots the scoring function (e.g. average RMSE over many 5-fold cross validation splits) as a function of the number of training data points used in the fit. This type of learning curve is helpful to assess whether the errors in your trained model may be sensitive to the size of the training data set, which may be particularly useful for small or sparse data sets. The second type of learning curve plots the model error as a function of the number of features in the feature matrix. This learning curve can directly inform feature selection and also assess if the model may be overfitting.

(VII) *Model form.* The choice of machine learning algorithm is often a central feature of constructing the best possible model. MAST-ML allows one to use nearly every type of regression and classification model algorithm available in scikit-learn. MAST-ML also allows construction of models which have more complicated input than simple numerical or string-based parameters (e.g., beyond just specifying a single kernel name and hyperparameter values for a kernel ridge regression model). For instance, MAST-ML allows the construction of custom kernels for use in Gaussian process regression, for example, a custom kernel which consists of multiplying a constant kernel with a radial basis function kernel. In addition, MAST-ML can import a previously trained scikit-learn model that one may want to either re-train or use in evaluating predictions on new data.

(VIII) *Model hyperparameter optimization.* Most machine learning models have one or more so-called hyperparameters, which are parameters that are not fit by the basic model fitting and need to be optimized separately in order to have the best possible model. For this task, MAST-ML currently leverages the grid search, randomized search, and Bayesian search methods contained in the scikit-learn and scikit-optimize (sometimes called skopt) packages.

(IX) *Model evaluation.* A central aspect of understanding the quality and domain of applicability of machine learning models hinges on model evaluation through some form of cross validation. Cross validation broadly consists of leaving out some portion of the data to serve as a validation sub data set, which is evaluated after a model is trained on the remaining training sub data set. In practice, there are many different methods one may use to construct training and validation sub data sets for cross validation. MAST-ML allows the use of cross validation techniques from scikit-learn as well as a number of custom routines. The available cross validation methods consist of random k-fold cross validation, leave one out cross validation (i.e., k=size of dataset), leave out group cross validation, leave out percent cross validation, leave out cluster cross validation, and leave out close composition cross validation. In leave out cluster cross validation, the data are first pre-clustered using a clustering algorithm, for example, K-means clustering, and each cluster is sequentially used as validation data analogously to leave out group cross validation. In leave out close composition cross validation, data with material compositions "close" to each other in composition space are placed into groups which are then analyzed analogously to leave out group cross validation. What constitutes as "close" can be specified by the user as a custom radial distance in composition space, where this radial distance is calculated from the atom fractions of the elements in the compositions that are near each other within a given L2 or L1 norm value. In addition, MAST-ML allows the user to perform a so-called "full fit" of the data, where all of the data is used in model training (i.e. no cross validation is performed), and this test can serve as a useful baseline for comparing to various cross validation tests.

(X) *Output Analysis.* Having output from an ML model easily analyzed to yield standard convenient statistics and visualizations is essential for enabling rapid progress and promoting best practices in model assessment. MAST-ML features a wide array of data analysis and plotting routines, which are implemented automatically during a run. Standard data plots include histograms of residuals, parity (true vs. predicted) plots showing the average values over all cross validation splits, the best and worst individual cross validation split (e.g. the best and worst folds in a 5-fold cross validation), the best and worst overall fitted result for every data point from cross validation, and, for leave out group cross validation tests, standard statistics like RMSE and $R^2$ are plotted for each group and as a function of group size, to allow one to easily assess predicative ability for specific left-out groups. MAST-ML also outputs a suite of standard error metrics for every data split (e.g. each fold of a 5-fold cross validation test) by default, for example the $R^2$ score, mean absolute error (MAE), root mean squared error (RMSE), and reduced error, which here is defined as the RMSE divided by the standard deviation of the dataset, among other metrics available through scikit-learn. Moreover, MAST-ML automatically outputs .csv files of all data used to make all analysis plots, as well as Jupyter notebook files which contain the relevant code snippet needed to reproduce a particular plot. The former feature is useful for identifying potential outlier data points in the analysis plots or performing further data analysis, and the latter feature allows one to directly reproduce and modify the generated plots to fine-tune details according to personal preference. All of these output data and plot files are generated for every split of cross validation conducted, as well as higher level summary plots averaged over all splits of a particular cross validation method. As the use of multiple models, cross validation routines, and even feature normalization routines can quickly result in a large number of individual tests and associated data, MAST-ML offers a high-level summary HTML document providing links and thumbnail images for each test performed.

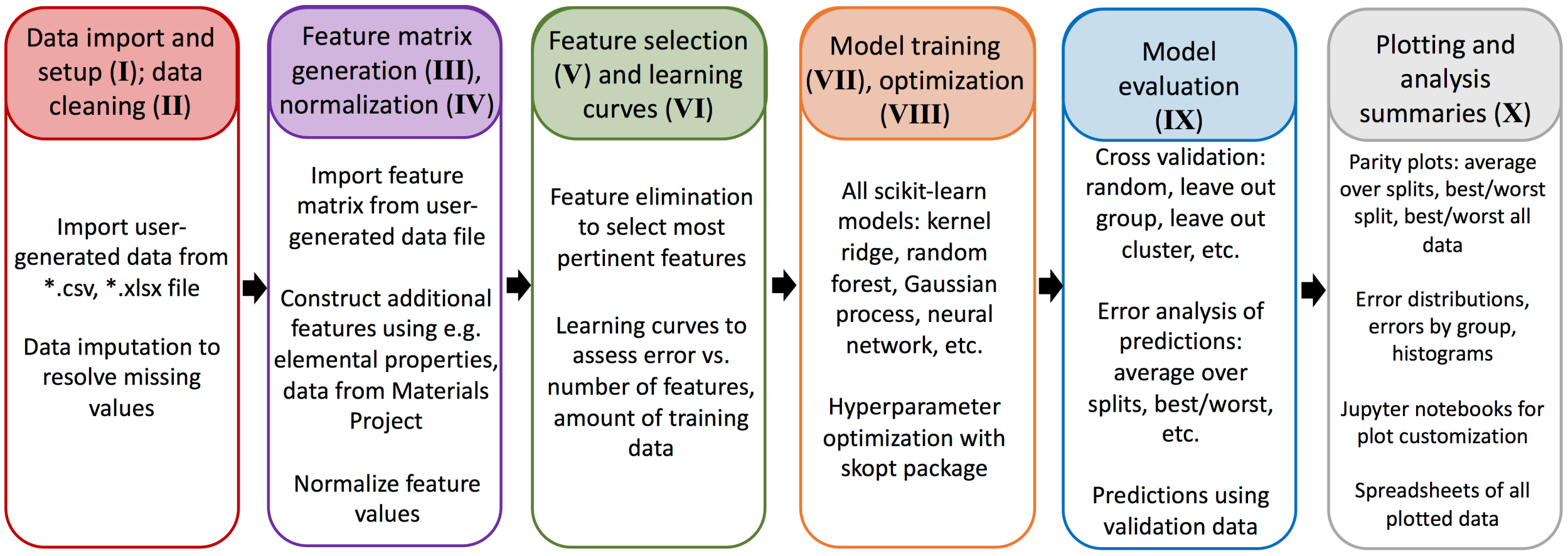

**Figure 1.** Overview of a typical supervised machine learning workflow. Each block represents a portion of the overall workflow. The roman numerals within each section of the workflow are meant to label components of the workflow that will be referenced in the main text and when discussing the MAST-ML input file structure in **Figure 3** and the structure of the MAST-ML output in **Figure 4**. The text in each block denotes operations one may conduct with MAST-ML as part of that portion of the workflow. Note that *MAST-ML will iterate through every combination of specified feature normalization, selection, machine learning model, and cross validation data split within a single run*.

Here, we would like to remark on the different ways MAST-ML can be used to obtain error bars on predicted data points. MAST-ML focuses on two widely used methods of generating error bars on predicted points: (1) through data resampling methods and (2) through predicted errors provided from specific machine learning models. Regarding predicted errors from data resampling methods, multiple iterations of standard cross validation (e.g. many iterations of 5-fold cross validation), leave out percent cross validation, and bootstrapping all provide stochastic methods of data resampling with replacement. Based on the calculated error (e.g. the root mean squared error) over all splits of these resampling routines, an average and standard deviation for each predicted data point can be obtained, and the standard deviations of the predictions of each data point function as an error bar on the predicted value of each data point. Regarding predicting errors provided from specific machine learning algorithms, MAST-ML currently supports error bar generation using Gaussian process regression and random forest regression models. For Gaussian process regression, the model provides a distribution of the predicted value for each data point, and we use the mean and standard deviation of the predicted distribution to assign a predicted value and error, respectively. For random forests, the error bars are obtained by taking the standard deviation of the predicted values for each decision tree estimator comprising the random forest model, which will result in an error bar on each data point. To help visualize some of these model error calculations, MAST-ML provides analysis plots of (cumulative) distributions of model residuals over (divided by) estimated model errors compared to (cumulative) normal distributions to assess model error estimates. This value of model residual divided by estimated model error is sometimes referred to as the *r*-statistic, and has been used to obtain valuable insight of predicted errors for random forests in the work of Ling et al.[60] and for Gaussian process regression in Lu et al.[41]

Finally, we note here that MAST-ML is continually under development, and examples of new features in development as of this writing include: new types of model statistical analysis, new types of analysis plots, new methods to perform cross-validation (e.g. different ways to automatically group input data), new model capabilities such as easy construction of deep neural networks with Keras, more informative calculations of predicted model errors (e.g. the jackknife-after-bootstrap method for random forest regression),[61] and more efficient methods to conduct nested cross validation with many sub data sets, among others. We expect that continual improvements and additional features will be added to MAST-ML in the future.

## 3. Obtaining and running MAST-ML:

MAST-ML may be obtained by downloading the source code from Github (https://github.com/uw-cmg/MAST-ML), downloading the code version accompanying this publication in the **Data Availability**, or performing a pip install via the command *pip install mastml*. If the source code is downloaded from Github or the **Data Availability**, please ensure the necessary dependencies are installed by running the accompanying setup.py file. More information on installing MAST-ML can be found in the online documentation (https://mastmldocs.readthedocs.io/en/latest/). Note that Python 3 is required to run MAST-ML properly, and it will work on both Mac and Windows machines. MAST-ML is distributed under an open source MIT license. MAST-ML requires two files to perform a run. The first file is a data file, which must be either a .csv or .xlsx file format, and will be described in detail in **Section 4**. The second file is the input file, which will be described in **Section 5**. The user may run MAST-ML in a couple different ways. The first way is to run the following Python module execution line in the Terminal/command line:

*python3 -m mastml.mastml_driver mastml/tests/conf/example_input.conf mastml/tests/csv/example_csv.csv -o results/mastml_testrun*

Here, *python3 -m mastml.mastml_driver* executes the *mastml_driver* module, *mastml/tests/conf/example_input.conf* provides the path to the input file with .conf extension (see **Section 5** for more information on the input file), *mastml/tests/csv/example_csv.csv* provides the path to the data file with .csv or .xlsx extension (see **Section 4** for more information on the data file), and *-o results/mastml_testrun* specifies the path to save the MAST-ML output. The second way to run MAST-ML is to call the *mastml.mastml_driver.main()* function via the Jupyter notebook contained in the MAST-ML repository (also available as part of the **Data Availability**). The Jupyter notebook simply needs to be updated to contain the proper paths providing the locations of the input file, data file, and results folder as described above. We note here that a version of MAST-ML is installed in the shared Python environment on Nanohub (accessible through the Jupyter notebook tool on Nanohub) and is freely available for use by anyone with a Nanohub account. Finally, the advanced user may also write their own Python script to import and call the *mastml.mastml_driver.main()* function as they see fit.

## 4. The MAST-ML data file

The data file contains all information on the target data *Y* to which a machine learning model is being fit and any input feature vectors *X* which form the feature matrix used to construct the fit. **Figure 2** shows a snippet of an example data file to show the recommended formatting. The data file can contain an arbitrary number of data instances (rows) and feature vectors (columns), though typical values are in the range of hundreds to thousands of data instances and tens to hundreds of feature vectors. All columns require names, as shown in **Figure 2**. The column names are necessary as MAST-ML uses pandas dataframes, which rely on column names for indexing and searching. It is also required that the target data *Y* vector be specified. In addition, the user can specify any number of *X* feature vectors, or start with no feature vectors if features will be generated within MAST-ML using one of the available feature generation schemes. Finally, any number of additional columns that are not part of the *X* or *Y* may also be specified (the columns marked as "Additional columns" in **Figure 2**). These additional columns can be used to preserve additional useful information on the dataset of interest, such as materials processing conditions, paper citations, or user notes. Note that for the purposes of illustrating the input/output structures in **Figure 2**, **Figure 3** and **Figure 4**, we have made use of an illustrative dataset of density functional theory (DFT) calculated diffusion activation energies of dilute solute diffusion in metal hosts from the work of work of Wu et al.[40] and Lu et al.[41] Note that the diffusion activation energies for solutes in each metal host are shifted by the diffusion activation energy of the host material, and this shifted diffusion activation energy can be equal to 0 (e.g. Al solute in an Al host) either positive or negative. The columns marked as "Additional columns" in **Figure 2** are optional and here specify (1) material compositions for the purpose of feature generation ("Material composition"), (2) group labels for leave out group cross validation test ("Host element"), (3) whether a data point is used as test data (i.e. never used in training or validation) ("predict_Pt"), and (4) a helpful note to the user ("Solute element"). Overall, these additional columns contain useful information but are not explicitly part of the feature matrix *X* or target data *Y* values. The user can easily specify which columns denote the features *X*, target data *Y*, and other miscellaneous information in the input file, which is described next in **Section 5**.

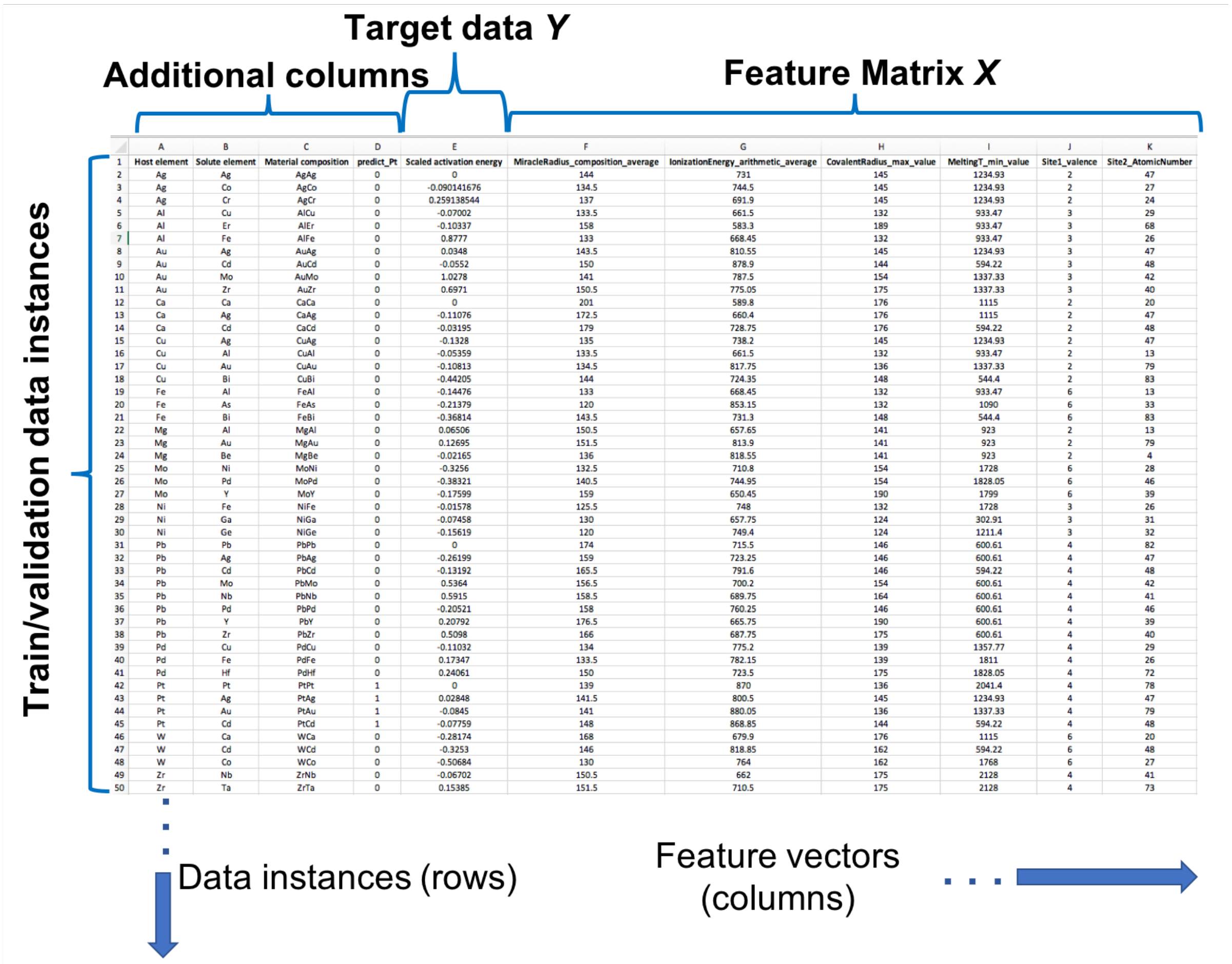

**Figure 2.** Example of the structure of the data (.csv or .xlsx) file used in MAST-ML. The data file consists of rows of individual data instances and columns of feature vectors. Every feature vector requires a name, as shown by the column headings under "Feature Matrix *X*". One column must denote the target data vector *Y*. The columns marked as "Additional columns" are optional and their meaning is described in the text, with more information on their use in **Section 5**. The data shown here is an illustrative subset of the dilute impurity activation barriers from the work of Wu et al.[40] and Lu et al.[41] This data file is included as part of the **Data Availability**.

## 5. The MAST-ML input file

The second file required to run MAST-ML is the input file, which is a text-based file that is both easily human- and machine-readable, and has the .conf file format. While the .conf file format is a general way to store system settings for Linux processes, this format was chosen here to meet the input file needs of the software package configobj,[62] which is used to parse the input file. **Figure 3** presents an example MAST-ML input file. In **Figure 3**, the input file is designed so that different section headings (first word blocks encased with "[ ]", e.g. "[DataCleaning]") closely correspond to a component of the supervised learning workflow described in **Section 2**, where the roman numerals are again used to label the different section(s) of the input file to denote the respective component of the supervised learning workflow from **Figure 1**. MAST-ML uses the configobj package,[62] which can easily parse an input .conf file and import it as a multilevel Python dictionary. The input file section headings (containing the "[ ]" character) and subsection headings (containing the "[[ ]]" character) and the associated parameters under these headings can be indented (or not) for organizational and ease of use purposes, as only the section and subsection heading characters determine the input file organization. Comments can be added with the "#" character (e.g., the first three lines of the example input file in **Figure 3**). For ease of use, numerous sections of the input file (FeatureNormalization, FeatureSelection, Models, HyperOpt, DataSplits) use subsection names that match class names contained in various scikit-learn modules. For example, in the Models section, the subsections "LinearRegression" and "KernelRidge", which denote different models, match the names of scikit-learn models contained in the sklearn.linear_model and sklearn.kernel_ridge modules, respectively. Further, the parameters of these models and all other cases where the (sub)sections match particular classes in scikit-learn (or other packages) are also identical to the parameters of their respective classes. For example, for the "KernelRidge" model, the parameter names "alpha", "gamma", and "kernel" match the parameters for the sklearn.kernel_ridge.KernelRidge class in scikit-learn. If specific values of these parameters are not specified in the input file by the user, then the scikit-learn (or respective package) default values are used.

One challenge in the workflow executed by MAST-ML is that the same parameters may be specified in multiple sections, and some examples of such parameter reuse are shown by the connected blue arrows crossing between sections in **Figure 3**. As a concrete example of parameter reuse across sections, all of the models the user specifies will be contained in the "[Models]" section. However, one may wish to use a specific model when conducting, e.g. feature selection and hyperparameter optimization. If that is desired, a particular model denoted in the "[Models]" section will also be listed in the "[FeatureSelection]" and "[HyperOpt]" sections. Note that the models from the "[Models]" section are listed just by their name in the other sections, not also with all associated parameters.

At the time of this writing, MAST-ML supports the use of multiple instances of the same model or cross validation data splitter type, and follows the convention of appending a unique string following the appropriate (sub)section name. For example, as shown in **Figure 3**, one can have multiple KernelRidge models, where the first instance is denoted as "KernelRidge", while the other models are denoted as "KernelRidge_select" and "KernelRidge_learn", which names were chosen (arbitrarily) because these models were used in the *FeatureSelection* and *LearningCurve* sections to evaluate the selection of pertinent features and construct learning curves, respectively. Note that models used in sections outside of the main *Models* section (like the above example of "KernelRidge_select" and "KernelRidge_learn" used in the *FeatureSelection* and *LearningCurve* sections, respectively) are subsequently removed from the list of models used in model training and testing, so that, in this case, multiple kernel ridge regression models with the same parameters are not used in model training and testing.

Here, we briefly describe the details of the MAST-ML run shown in **Figure 3**, going section-by-section in this example input file. In the list below, the section headings are given as *"Input file heading"* (*"Supervised learning workflow heading"*), which highlights the direct connection between each section of the MAST-ML input file and each component of the general supervised machine learning workflow discussed in **Section 2** and shown in **Figure 1**. A more thorough description of the MAST-ML input file, along with a list of available parameter values for each field, and step-by-step tutorials and example files, can be found online as part of the MAST-ML documentation (https://mastmldocs.readthedocs.io/en/latest/). In some cases, the available input file parameters are not listed exhaustively in the online MAST-ML documentation because the available values come directly from separately documented scikit-learn methods. In these cases, the reader is directed to the appropriate portion of the scikit-learn online documentation. The online MAST-ML documentation will also be continuously updated as new features become available in the future.

(I) *GeneralSetup (Data import)*. The GeneralSetup section allows the user to control the process of data import and MAST-ML run setup. There are seven main parameters for this section, which we will describe in some detail here as they are central to successfully conducting a MAST-ML run:

a. *input_features*: This parameter controls which columns of the data file are considered as part of the feature matrix *X*. The user can specify each column individually, or use the "Auto" keyword to assume all columns that are not designated in the *input_target* or *input_other* fields are part of the feature matrix.

b. *input_target*: This parameter controls which column of the data file denote the target *Y* data. Note that only a single target value *Y* may be specified, and to examine other *Y* values requires separate MAST-ML runs.

c. *input_other*: This parameter is used to denote additional columns present in the data file which are not part of the target data *Y* or feature matrix *X*. From **Figure 2**, and as shown in **Figure 3**, for this example the columns "Host element", "Solute element", "Material composition", and "predict_Pt" are all extra columns.

d. *input_grouping*: This parameter denotes any columns that are used to provide group labels which bin the data into separate groups for use in a leave out group cross validation analysis. For this example, "Host element" is used as a grouping feature.

e. *input_test*: This parameter is used to denote any columns which label particular data points in the data file as test data, i.e., data only used in model predictions and never used in training or validation during cross validation. Note that a particular data point should be labeled with a "0" if not used as test data and "1" if it is to be used as test data. For this example, "predict_Pt" is used to label data where Pt is the host (see **Figure 2**) and is used as test data.

f. *randomizer*: This feature provides the user the option to randomly shuffle the *X* feature matrix rows. The point of the randomizer is to provide a comparative "null" test where any potential physical or chemical relationship between the feature matrix and the target data is removed. This is useful to ascertain the robustness of correlations and the amount of physical insight (as opposed to just fitting numerical noise) present in the correlation between the feature matrix and the target data.

g. *metrics*: This parameter controls which model evaluation statistics are evaluated and reported on analysis plots (e.g., mean absolute error, $R^2$ score, etc.). One can set this parameter to use a canonical set of evaluation statistics with the "Auto" keyword.

(II) *DataCleaning (Data cleaning)*. The *DataCleaning* section of the input file allows the user to clean their data to remove rows or columns that contain empty or not-a-number (NaN) fields, or fill in these fields using imputation or principal component analysis methods. In this example:

a. *cleaning_method:* (remove, imputation, ppca). Method used to perform the data cleaning. In this example, missing data values are filled in using imputation.

b. *imputation_strategy:* (mean, median, mode). For imputation only, how the imputation is performed. In this example, the mean value of the feature vector is used as the missing value.

(III) *FeatureGeneration (Feature generation)*. The *FeatureGeneration* section provides the ability to construct additional feature vectors based on elemental properties, queries to the materials databases Citrination and the Materials Project, and the construction of structural fingerprints based on routines from the matminer package. In this example, the MAGPIE method is being used to construct a set of elemental property features.

a. *composition_feature*: This is the name of the appropriate column of the data file (in this case, "Material composition", see **Figure 2**) which specifies the material compositions of each data point with which to build the set of elemental property features. The composition strings must be interpretable as pymatgen Composition objects, e.g. "La0.75Sr0.25MnO3". In addition, the user may specify a composition using square brackets to denote different sublattices, e.g. "[La0.75Sr0.25][Mn][O3]". Doing so will result in feature generation conducted on a per-sublattice basis instead of for the entire material composition.

b. *feature_types*: (composition_avg, arithmetic_avg, max, min, difference, elements). The types of elemental features to construct.

(IV) *FeatureNormalization (Feature normalization)*. The *FeatureNormalization* section enables one to scale the input or generated features using well-known feature normalization algorithms available in scikit-learn. In this example, the scikit-learn *StandardScaler* method is used with default values, which scale the data to have mean of zero and standard deviation of one.

(V) *FeatureSelection (Feature selection)*. This section provides the means to select the set of most pertinent features from the full feature matrix. In this case, the *SequentialForwardSelector* routine from the mlxtend package is used:

a. *estimator*: Used to specify a particular machine learning model to conduct feature selection. Here, the name *KernelRidge_select* is specified, which is a kernel ridge regression model that must also be specified in the *Models* section.

b. *k_features*: The number of features to select.

(VI) *LearningCurve (Learning curve)*. This section controls how to formulate learning curves to assess the optimal number of features to use and to evaluate model performance vs. training set size.

a. *estimator*: Used to specify a particular machine learning model to conduct feature selection. Here, the name *KernelRidge_learn* is specified, which is a kernel ridge regression model that must also be specified in the *Models* section.

b. *cv*: The type of cross validation data splitter to use to evaluate the model score. Here, the name *RepeatedKFold_learn* is specified, which performs multiple iterations of leave out K-fold cross validation (specifically, 2 iterations of leave out 5-fold cross validation) that must also be specified in the *DataSplits* section.

c. *scoring*: The metric used to assess model performance. In this case, the root mean squared error is used.

d. *n_features_to_select*: The number of features to select.

e. *selector_name*: The method of feature selection used to evaluate each step of the learning curve. Must be a valid routine one would use in the *FeatureSelection* section.

(VII) *Models (Model training)*. This section controls which models are used to evaluate every specified cross validation data splitting test (see *DataSplits* section). In addition, other models may be specified here which are used in other components of the workflow, such as *KernelRidge_select* and *KernelRidge_learn* which were used in the *FeatureSelection* and *LearningCurve* sections, respectively. For this MAST-ML run, a linear regression (*LinearRegression*), random forest model (*RandomForestRegressor*), kernel ridge regression (*KernelRidge*), and previously fit random forest model (*ModelImport*) will be evaluated. The *ModelImport* subsection allows users to import previously fit models (as a .pkl file) by specifying the path where the model is saved via the *model_path* parameter. Note that MAST-ML will automatically save all fit models (for every train/test split) as a .pkl file.

(VIII) *HyperOpt (Model optimization)*. This section is used to specify methods to optimize the hyperparameters of a particular model. Currently, hyperparameter optimization using a grid search (*GridSearch*), randomized search (*RandomizedSearch*), and Bayesian search (*BayesianSearch*) are available.

a. *estimator*: Used to specify a particular machine learning model to conduct hyperparameter optimization. Here, the name *KernelRidge* is specified, which is a kernel ridge regression model that must also be specified in the *Models* section.

b. *cv*: The type of cross validation data splitter to use to evaluate the model score. Here, the name *RepeatedKFold* is specified, which performs multiple iterations of leave out k-fold cross validation (specifically, 2 iterations of leave out 5-fold cross validation) that must also be specified in the *DataSplits* section.

c. *param_names*: The variable names of the hyperparameters to be optimized. For the case of a *KernelRidge* model, these parameters which were chosen to optimize are *alpha* and *gamma*. Note that the parameter names need to be delimited with a semicolon.

*d.* *param_values*: This field is used to specify the grid of values to use in optimization. The notation follows the pattern *min_value max_value number_of_points spacing_type data_type*, where *min_value* is the minimum grid value, *max_value* is the maximum grid value, *number_of_points* is the number of parameter grid values to evaluate, *spacing_type* denotes linear or logarithmic spacing ("*lin*" or "*log*", respectively), and the *data_type* is used to denote whether the gridded values are integers or floats ("*int*" or "*float*", respectively). As with the *param_names* field, the *param_values* for different parameters must also be delimited with a semicolon. Note that the method to specify the space of values to evaluate differs slightly based on which hyperparameter optimization routine is used. Consult the MAST-ML online documentation for examples of how to properly set up *param_values* for *GridSearch*, *RandomizedSearch* and *BayesianSearch* methods.

e. *scoring*: The metric used to assess model performance. In this case, the root mean squared error is used.

(IX) *DataSplits (Model evaluation)*. This section is used to specify the different types of cross validation splits used to evaluate each machine learning model. In this example, the different specified splits are *NoSplit*, *RepeatedKFold*, *RepeatedKFold_learn*, and *LeaveOneGroupOut*. The *NoSplit* case is used as a point of comparison, as all data is used in training and testing (*i.e.*, no actual cross validation). *RepeatedKFold* conducts random leave out cross validation (in this case, 2 iterations of leave-out 20%) and *RepeatedKFold_learn* does the same thing but is listed here as it is used to construct the learning curve (see *LearningCurve* section). Lastly, *LeaveOneGroupOut* conducts a leave-out group cross validation routine, where the groups are manually specified by the user in their data file (specifically, the "Host element" column of the data file as shown in **Figure 2**) and listed here with the parameter *grouping_column*. The *grouping_column* value must match the corresponding column name in the data file.

(X) *MiscSettings (Plotting and analysis settings)*. This section is used to specify whether certain types of output analysis plots are saved to the MAST-ML run results directory.

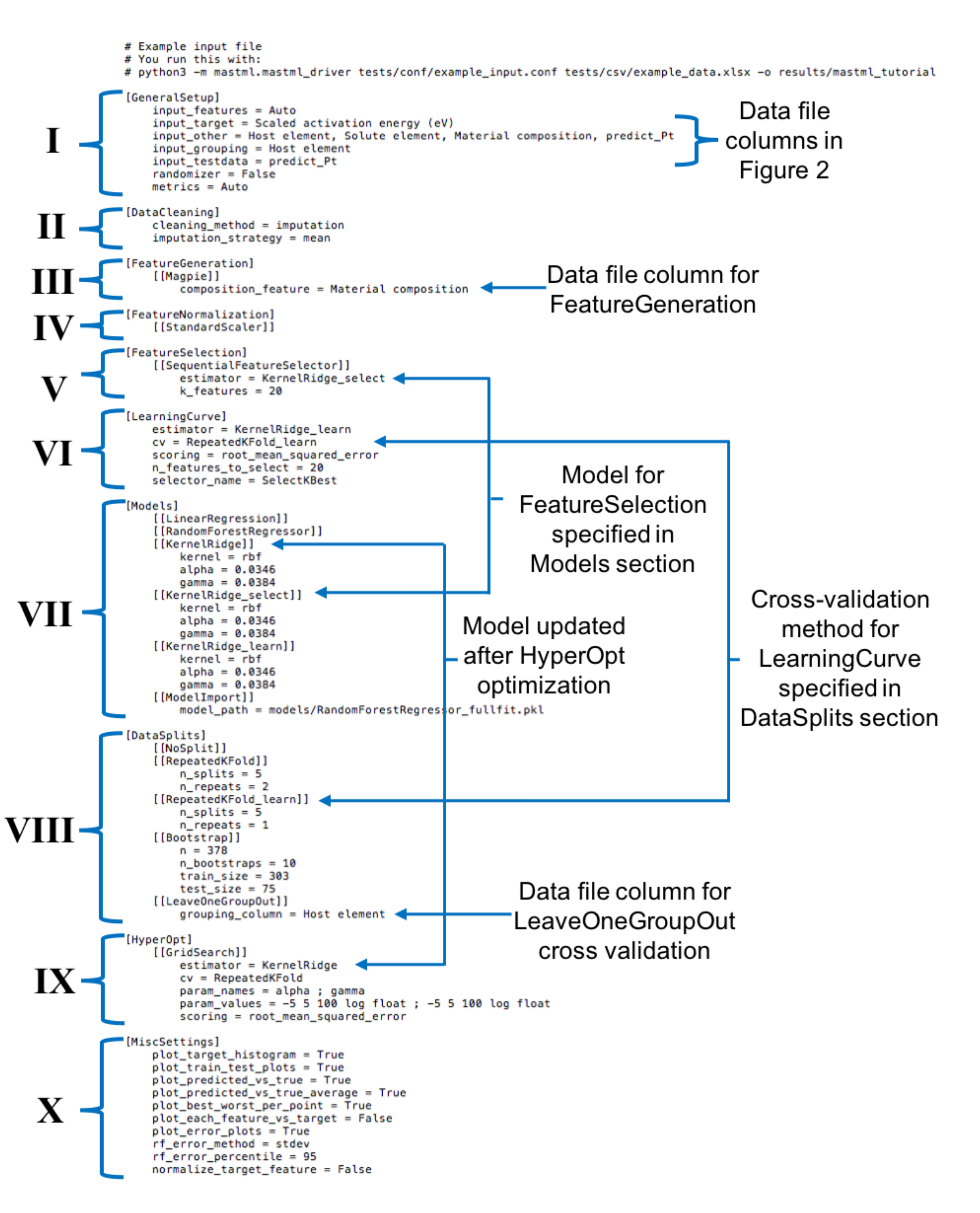

**Figure 3.** Example of the MAST-ML input file. The roman numeral labels for each section block correspond to the equivalent sections of the supervised machine learning workflow shown in **Figure 1** and discussed in **Section 2**. Section headings are denoted with the "[ ]" character and subsection headings are denoted with the "[[ ]]" character. Indentations can be used to easily organize each section and comments can be included with the "#" character. This input file is included as part of the **Data Availability**.

## 6. MAST-ML output structure and contents:

Here, we provide an overview of how the output of a MAST-ML run is organized, and the types of output available to the user. **Figure 4** contains an overview of the resulting output directory tree for the MAST-ML run conducted using the data file shown in **Figure 2** and the input file shown in **Figure 3**. In **Figure 4**, the roman numerals are again used to label different components of the supervised learning workflow outlined in **Figure 1** and labeled as sections of the input file in **Figure 3**. Here, these different steps of the supervised machine learning workflow are separated as different levels of the output directory tree. The user can specify the path to save all MAST-ML run results. Here, the *results* folder within the main MAST-ML directory was chosen. Note that one can re-use the same output folder path, and MAST-ML will automatically append datetime data to create a unique folder path. The output directory tree shown in **Figure 4** was generated using the data file shown in **Figure 2** and the input file from **Figure 3**. These files are available as part of the **Data Availability**. Due to the large volume and variety of files generated from a MAST-ML run, an exhaustive list and description of each type of output file will not be provided here. However, to give the prospective user a general idea of the type of output MAST-ML provides, here we list the main categories of output files along with example instances of each file type. The main types of output files consist of:

a.) Image files (.png): These files typically contain data analysis plots saved in .png format. These data analysis plots consist of data histograms, parity scatter plots, error distribution plots, general scatter plots of feature vs. target values, and learning curves. These plots are saved at various stages of the MAST-ML workflow, from every train/test split to average summaries for each cross validation data split. A set of example plots is shown in **Figure 5** and will be discussed more in the following paragraph. Note that the code used to generate these figures is also output as a Jupyter notebook (see "Jupyter Notebooks" below) to allow straightforward modifications.

b.) Data files (.csv): These files contain data from either various stages of modifying the input data file or from analysis output. For example, the data file provided for a MAST-ML run is copied to the save directory and modified versions of that file are saved after data cleaning, feature generation, feature normalization, and feature selection routines have completed. In addition, a file containing a high-level summary of input data statistics is provided (for example, showing the distribution of the target *Y* data, its median, average value, etc.). Finally, .csv files are provided which contain the data used to construct all of the analysis data plots to assist the user in identifying problematic data points, performing further analysis or creating new types of analysis plots specific to their research project. In general, for each plotted result, an image file, .csv data file, and customizable Jupyter notebook are generated.

c.) Model files (.pkl): The saved model objects (e.g. a trained *KernelRidge* model) are saved as .pkl files for every train/test split for each cross validation data splitter.

d.) Run statistics summary (.txt): A high-level summary of the average train, validation, and test (if applicable) statistics is provided (e.g. the average of the mean absolute error of fits of training data, validation data, and test data (if applicable) over all cross validation splits).

e.) Jupyter notebooks (.ipynb): Jupyter notebooks containing source code used to make each type of data analysis plot image discussed above are provided to assist the user in making small modifications to saved plots based on personal preference.

f.) Log file (.log): Python log file which contains basic run information and runtime warnings and errors for the MAST-ML run.

g.) Summary HTML document (.html): An HTML document which shows, at a high level, the sets of analysis plots and data files contained in the MAST-ML results output directory tree, as well as links to specific folders containing the respective plots/data files shown.

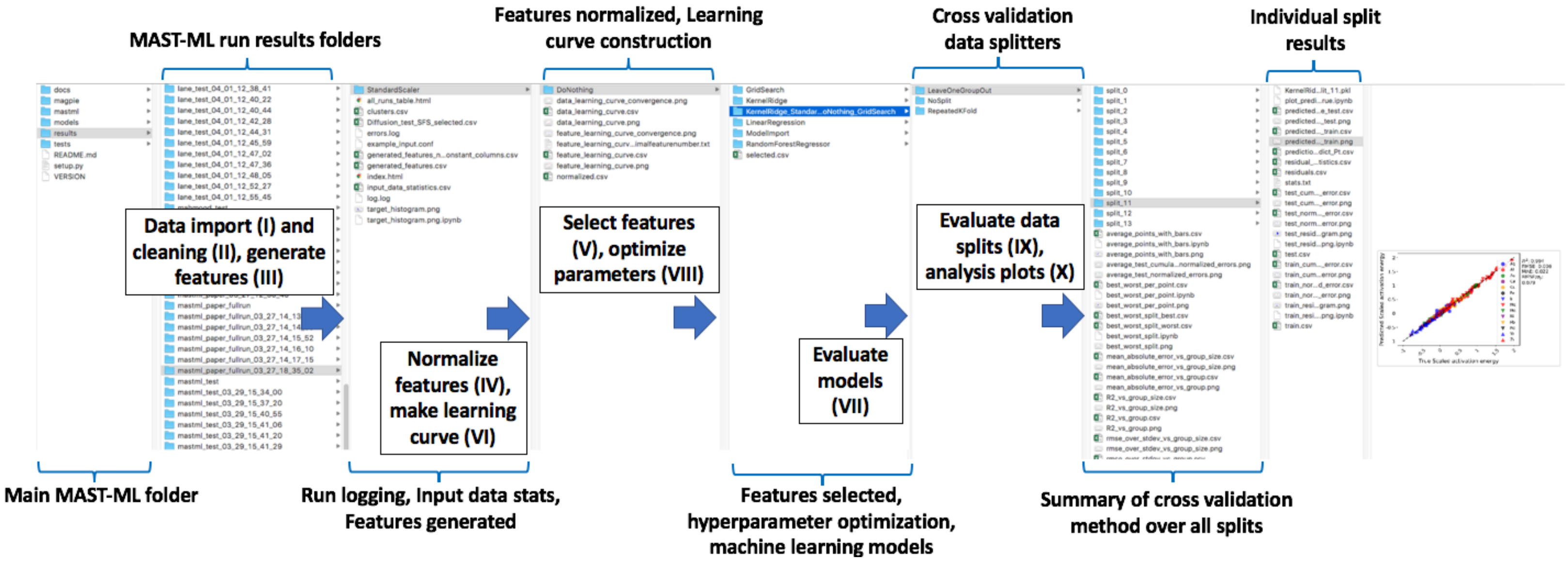

**Figure 4.** Example MAST-ML output directory tree structure and associated contents. The roman numeral labels correspond to the equivalent sections of the supervised machine learning workflow shown in **Figure 1** and the input file sections shown in **Figure 3**.

**Figure 5** provides an example overview of the different types of data analysis plots provided from a MAST-ML run. These plots were obtained from the output run shown in **Figure 4**, generated from running MAST-ML on the data file in **Figure 2** and the input file in **Figure 3**. These plots were automatically generated in the MAST-ML run and were not modified in any way for display in this publication after being produced in MAST-ML, except to move legend entries to provide an improved presentation of many images grouped together in a single figure. We note here that these example plots in **Figure 5** are meant purely as an illustrative example to showcase features of MAST-ML and are not meant to be a robust scientific result of machine learning predictions of dilute solute diffusion activation energies in metals. More details of the key results of recent studies of machine learning on impurity diffusion and other recent studies using MAST-ML for other machine learning problems in materials science are provided in **Section 7** and can be found in the work of Wu et al.[40] and Lu et al.[41] **Figure 5** showcases numerous analysis plot types, with the first being an example histogram of input data **Figure 5A**. This histogram provides basic statistics of target *Y* data, such as mean, median, etc. Shown in **Figure 5B** is a learning curve plotting the root mean squared error as a function of amount of training data, constructed using a kernel ridge regression model and 5-fold cross validation. In **Figure 5B**, the red (blue) data are the root mean square errors of the validation (train) data, respectively, and the shaded area represents the standard deviation in the root mean squared error calculated over all random leave-out cross validation splits. In **Figure 5C**, a parity plot of predicted vs. true values of the scaled diffusion barrier activation energy is shown, where the data are average validation data values over two iterations of random 5-fold cross validation. The error bars on the points represent standard deviations of the root mean squared error over all random 5-fold cross validation splits. In **Figure 5D**, a parity plot showing predicted vs. true values of the scaled diffusion barrier activation energy is shown, where the data are validation data values collected from all groups in leave out group cross validation, where the model is trained on all data except for Pt (recall for this example Pt is left out as a test data set). In **Figure 5E**, a parity plot showing the same data as in **Figure 5C**, except the best (red) and worst (blue) split results are shown instead of the average over all splits. The plot in **Figure 5F** shows a further analysis of a leave-out group cross validation test, where the mean absolute error of the test data is shown as a function of group label, demonstrating how well particular groups are predicted relative to others. **Figure 5G** shows a plot of error distributions comparing an analytical Gaussian curve (blue), model residual (i.e. predicted minus true) values (green), and the model errors resulting from a random forest model (purple). The $x$-axis is the ratio of the residual value to the standard deviation of the predicted model error, which is calculated for every data point as described in **Section 2**. Finally, the plot in **Figure 5H** is the same data as shown in **Figure 5G**, but as a cumulative error distribution.

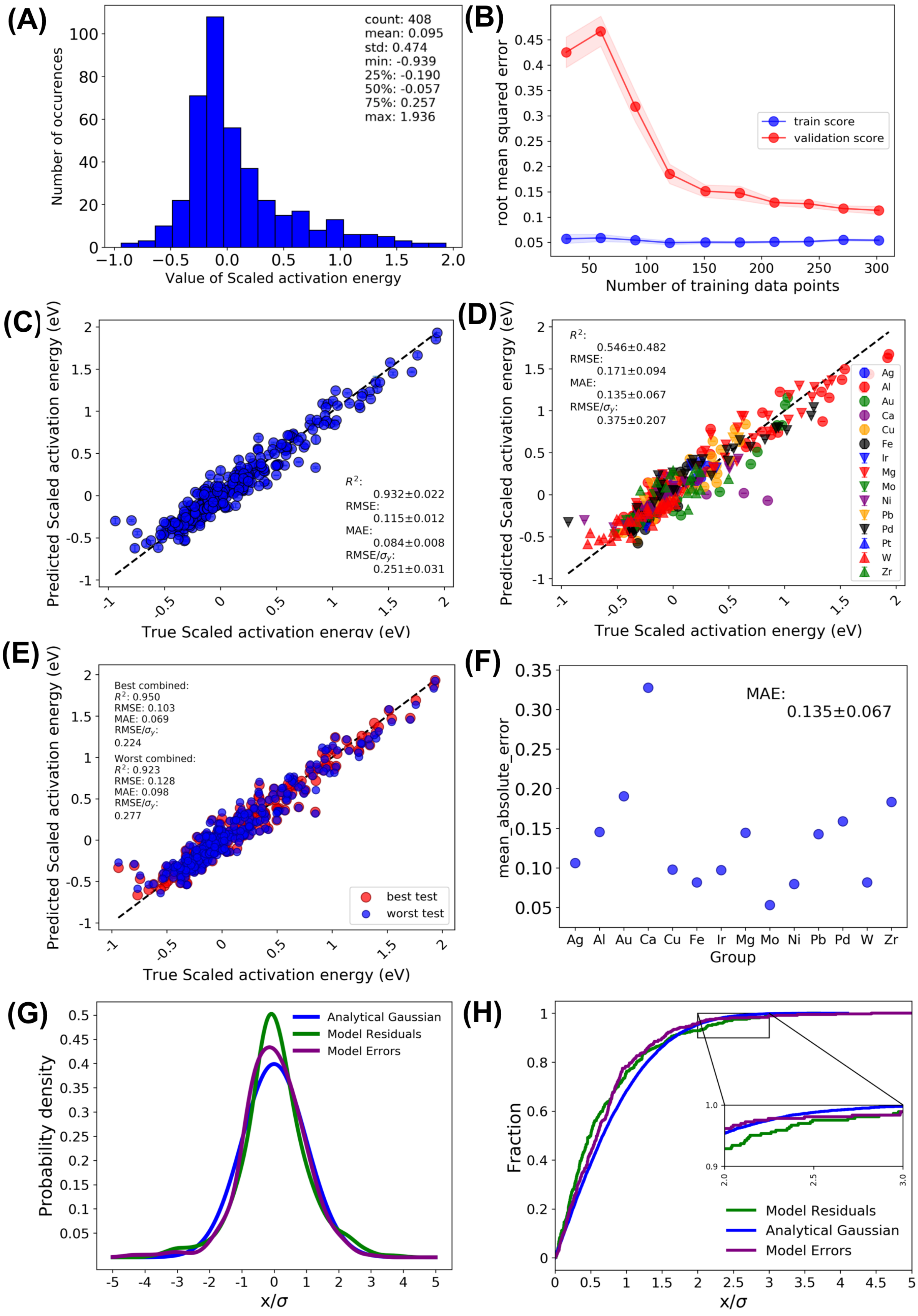
(A)
count: 408
mean: 0.095
std: 0.474
min: -0.939
25%: -0.190
50%: -0.057
75%: 0.257
max: 1.936
Number of occurences
Value of Scaled activation energy
(B)
root mean squared error
train score
validation score
Number of training data points
(C)
Predicted Scaled activation energy (eV)
True Scaled activation energy (eV)
$R^2$: 0.932±0.022
RMSE: 0.115±0.012
MAE: 0.084±0.008
RMSE/$\sigma_y$: 0.251±0.031
(D)
$R^2$: 0.546±0.482
RMSE: 0.171±0.094
MAE: 0.135±0.067
RMSE/$\sigma_y$: 0.375±0.207
Ag Al Au Ca Cu Fe Ir Mg Mo Ni Pb Pd Pt W Zr
Predicted Scaled activation energy (eV)
True Scaled activation energy (eV)
(E)
Best combined:
$R^2$: 0.950
RMSE: 0.103
MAE: 0.069
RMSE/$\sigma_y$: 0.224
Worst combined:
$R^2$: 0.923
RMSE: 0.128
MAE: 0.098
RMSE/$\sigma_y$: 0.277
best test
worst test
Predicted Scaled activation energy (eV)
True Scaled activation energy (eV)
(F)
MAE: 0.135±0.067
mean_absolute_error
Ag Al Au Ca Cu Fe Ir Mg Mo Ni Pb Pd W Zr
Group
(G)
Analytical Gaussian
Model Residuals
Model Errors
Probability density
x/σ
(H)
Model Residuals
Analytical Gaussian
Model Errors
Fraction
x/σ

**Figure 5.** Examples of data analysis plots provided as output from the example MAST-ML run discussed throughout this work. (A) Histogram of input data providing basic statistics of target *Y* data. (B) Learning curve plotting the root mean squared error as a function of amount of training data, constructed using a kernel ridge regression model and random 5-fold cross validation. (C) Parity plot showing predicted vs. true values of the scaled diffusion barrier activation energy, where the data are average validation data values over two iterations of random 5-fold cross validation. (D) Parity plot showing predicted vs. true values of the scaled diffusion barrier activation energy, where the data are all validation data values in a leave out group cross validation split. (E) Parity plot showing the same data as in (C), except displaying the best and worst split results instead of the average over all splits. (F) The validation data mean absolute error as a function of each group in a leave out group cross validation test. (G) Error distributions showing comparison of an analytical Gaussian curve (blue), model residual (i.e. predicted minus true) values (green), and the model errors resulting from a random forest model (purple). The *x*-axis is the ratio of the residual value to the standard deviation of the predicted model error, which is calculated for every data point. (H) The same data as shown in (G), but as a cumulative error distribution. See text for more details.

## 7. Recent materials research performed with MAST-ML

From early 2018 until the time of this publication (about a year and a half as of this writing), MAST-ML has been used at various stages of development on a number of materials science studies utilizing different scientific datasets for a range of materials applications. These studies include:

a) *Perovskite oxide stability*: The work of Li et al.[32] investigated the thermodynamic stability of nearly 2000 perovskite oxides calculated using DFT. Predicting and understanding the stability of perovskite oxides is a key aspect of engineering improved oxygen-active materials such as cathodes for solid oxide fuel cells. Li et al. found that a Gaussian kernel ridge regression model could predict the stability of perovskite oxides (represented as the energy above the convex hull) with an average root mean square error of about 29 meV/atom from random leave-out cross validation, close to DFT accuracy.
b) *Dilute solute diffusion in metals*: The work of Wu et al.[40] and Lu et al.[41] examined the activation energy for dilute solute diffusion in pure metal hosts using Gaussian kernel ridge regression (Wu et al. and Lu et al.) and Gaussian process regression (Lu et al.). Understanding the diffusivity of solutes in metal hosts is important for metal alloy engineering, such as controlling the effects of precipitation hardening and high temperature creep. Wu et al. obtained a root mean square error between 0.2-0.25 eV (depending on the host being predicted) for leave-out-group cross validation using a database of 218 activation energies while Lu et al. obtained an average root mean square error of just 0.15 eV for leave-out-group cross validation averaged over all groups, both values obtained using kernel ridge regression models. The improved predictions of Lu et al. were the result of using an expanded database of 408 activation energies and improved methods of feature selection and hyperparameter optimization incorporating leave-out-group cross validation tests.

c) *Effective charge in metals for electromigration:* The study from Liu et al.[39] detailed results of predicting the effective charge of an array of metals and dilute alloys. The effective charge is the main quantity governing the electromigration effect, and understanding and controlling electromigration is key to the successful use of metal alloys as interconnects in integrated circuits. Liu et al. found using multivariate linear models that a number of transition and noble metals (e.g. Ag, Au, Cu, Pt) doped in Co (a suggested replacement for Cu as an interconnect material) should result in very little electromigration, offering useful design avenues for new electronic interconnect alloys.

The studies described above are a demonstration of the power of software packages like MAST-ML and those complementary to it to accelerate the use of machine learning methods in materials research. These studies also show how MAST-ML can be used to generate highly informative machine learning models that can be used to improve materials engineering and inform materials discovery in a diverse array of applications and design spaces.

## 8. Summary and future outlook

The increasingly widespread adoption of materials informatics in the greater materials research community requires the access of easy-to-use, open access software tools which can provide quality research results without a large amount of experience necessary for their use. The current paper introduces the Materials Simulation Toolkit for Machine Learning (MAST-ML), an open access Python package which seeks to accelerate and codify the use of machine learning in materials science (i.e. materials informatics) and lower the barrier for entry to conducting sophisticated supervised learning problems. Further, we see MAST-ML as part of an ecosystem of open source software contributions in the field of materials informatics. While MAST-ML currently leverages key functionality of some existing packages like matminer, MAST-ML is constantly under development, with both new features unique only to MAST-ML and broader integration with other existing and emerging software packages planned for the future. Overall, it is our desire to design tools that enable acceleration of innovative data-driven materials research. We are therefore keen on addressing the needs of the greater materials informatics community. We are open to suggestions and collaborations of improving and expanding the MAST-ML code to incorporate new model types, new statistical tests, new workflows, and other desired features which would strengthen and expand materials informatics as an impactful research enterprise.

**Acknowledgements**

The authors gratefully acknowledge funding provided by the NSF Software Infrastructure for Sustained Innovation (SI2) award No. 1148011 and the NSF DMREF award number DMR-1332851. The NSF SI2 award No. 1148011 funded Ryan Jacobs, Tam Mayeshiba, Luke Miles, Max Williams and Matthew Turner. The NSF DMREF award number DMR-1332851 funded Ben Afflerbach. The authors wish to thank all of those who contributed to the development of MAST-ML in various ways, such as through (1) commits to the MAST-ML source code on Github (https://github.com/uw-cmg/MAST-ML), (2) contributed analysis scripts which were integrated to become MAST-ML features, (3) useful discussions with the authors of this paper on how to improve MAST-ML, (4) testing of different MAST-ML features to eliminate bugs and streamline performance (listed in alphabetical order): Alex Do, Shuo Han, Wei Li, Dr. Yu-chen Liu, Haijin Lu, Vanessa Meschke, Dr. Maciej Polak, Alex Politowicz, Sam Wagner, Zuf Wang, Dr. Logan Ward, Kangqi Xi, and Linda Xiao.

**CRediT authorship contribution statement**

**Ryan Jacobs:** Investigation, software, writing – original draft, writing – review and editing. **Tam Mayeshiba:** software, writing – review and editing, validation. **Ben Afflerbach:** Software, validation, writing – review and editing. **Luke Miles:** Software, validation. **Max Williams:** Software, validation. **Matthew Turner:** Software, validation. **Rapheal Finkel:** Conceptualization, supervision, project administration, writing – review and editing. **Dane Morgan:** Conceptualization, supervision, project administration, writing – review and editing.

**Data availability**

The raw data required to reproduce these findings are available to download from https://doi.org//10.6084/m9.figshare.7418492. The processed data required to reproduce these findings are available to download from https://doi.org/10.6084/m9.figshare.9977501.v1. The most up-to-date version of MAST-ML can be found on Github (https://github.com/uw-cmg/MAST-ML). The code and usage documentation (including tutorials) for MAST-ML can be found on Github and ReadtheDocs: https://mastmldocs.readthedocs.io/en/latest/index.html.

## References

[1] T. Kailil, C. Wadia, Materials Genome Initiative for Global Competitiveness, Https://www.mgi.gov/sites/default/files/documents/materials_genome_initiative-Final.pdf. (2011) 1–18. doi:10.1021/cen-09131-govpol1.

[2] J.J. de Pablo, N.E. Jackson, M.A. Webb, L.-Q. Chen, J.E. Moore, D. Morgan, R. Jacobs, T. Pollock, D.G. Schlom, E.S. Toberer, J. Analytis, I. Dabo, D.M. DeLongchamp, G.A. Fiete, G.M. Grason, G. Hautier, Y. Mo, K. Rajan, E.J. Reed, E. Rodriguez, V. Stevanovic, J. Suntivich, K. Thornton, J.-C. Zhao, New frontiers for the materials genome initiative, Npj Comput. Mater. 5 (2019).

[3] S. Curtarolo, W. Setyawan, G.L.W. Hart, M. Jahnatek, R. V. Chepulskii, R.H. Taylor, S. Wang, J. Xue, K. Yang, O. Levy, M.J. Mehl, H.T. Stokes, D.O. Demchenko, D. Morgan, AFLOW: An automatic framework for high-throughput materials discovery, Comput. Mater. Sci. 58 (2012) 218–226.

[4] A. Jain, S.P. Ong, G. Hautier, W. Chen, W.D. Richards, S. Dacek, S. Cholia, D. Gunter, D. Skinner, G. Ceder, K.A. Persson, Commentary : The Materials Project : A materials genome approach to accelerating materials innovation, APL Mater. 1 (2013) 11002.

[5] J.E. Saal, S. Kirklin, M. Aykol, B. Meredig, C. Wolverton, Materials design and discovery with high-throughput density functional theory: The open quantum materials database (OQMD), JOM. 65 (2013) 1501–1509.

[6] B. Blaiszik, K. Chard, J. Pruyne, R. Ananthakrishnan, S. Tuecke, I. Foster, The Materials Data Facility: Data Services to Advance Materials Science Research, JOM. 68 (2016) 2045–2052.

[7] D.D. Landis, J.S. Hummelshøj, S. Nestorov, J. Greeley, M. Dułlak, T. Bligaard, J.K. Nørskov, K.W. Jacobsen, The computational materials repository, Comput. Sci. Eng. 14 (2012) 51–57.

[8] J. O'Mara, B. Meredig, K. Michel, Materials Data Infrastructure: A Case Study of the

Citrination Platform to Examine Data Import, Storage, and Access, JOM. 68 (2016) 2031–2034.

[9] A. Agrawal, A. Choudhary, Perspective: Materials informatics and big data: Realization of the “fourth paradigm” of science in materials science, APL Mater. 4 (2016).

[10] R. Ramprasad, R. Batra, G. Pilania, A. Mannodi-Kanakkithodi, C. Kim, Machine learning in materials informatics: Recent applications and prospects, Npj Comput. Mater. 3 (2017).

[11] D.M. Dimiduk, E.A. Holm, S.R. Niezgoda, Perspectives on the Impact of Machine Learning, Deep Learning, and Artificial Intelligence on Materials, Processes, and Structures Engineering, Integr. Mater. Manuf. Innov. 7 (2018) 157–172.

[12] D. Morgan, G. Ceder, Data Mining in Materials Development, Handb. Mater. Model. (2005) 395–421.

[13] T. Mueller, A.G. Kusne, R. Ramprasad, Machine learning in materials science: Recent progress and emerging applications, Rev. Comput. Chem. 29 (2016).

[14] K.T. Butler, D.W. Davies, H. Cartwright, O. Isayev, A. Walsh, Machine learning for molecular and materials science, Nature. 559 (2018) 547–555.

[15] P. Ball, Using artificial intelligence to accelerate materials development, MRS Bull. (2019) 335–344.

[16] A.L. Ferguson, Machine learning and data science in soft materials engineering, J. Phys. Condens. Matter. (2018).

[17] B.L. DeCost, T. Francis, E.A. Holm, Exploring the microstructure manifold: Image texture representations applied to ultrahigh carbon steel microstructures, Acta Mater. 133 (2017) 30–40.

[18] V. Stanev, V. Vesselinov, A.G. Kusne, G. Antoszewski, I. Takeuchi, B. Alexandrov, Unsupervised Phase Mapping of X-ray Diffraction Data by Nonnegative Matrix Factorization Integrated with Custom Clustering, Npj Comput. Mater. (2018).

[19] N. Lubbers, T. Lookman, K. Barros, Inferring low-dimensional microstructure representations using convolutional neural networks, Phys. Rev. E. 96 (2017) 52111.

[20] W. Ye, C. Chen, Z. Wang, I. Chu, S.P. Ong, Deep neural networks for accurate predictions of crystal stability, Nat. Commun. (2018) 1–6.

[21] D. Jha, L. Ward, A. Paul, W. Liao, A. Choudhary, ElemNet : Deep Learning the Chemistry of Materials From Only Elemental Composition, Sci. Rep. 8 (2018) 1–13.

[22] W. Li, K.G. Field, D. Morgan, Automated defect analysis in electron microscopic images, Npj Comput. Mater. 4 (2018) 1–9.

[23] J.R. Hattrick-Simpers, J.M. Gregoire, A.G. Kusne, Perspective: Composition-structure-property mapping in high-throughput experiments: Turning data into knowledge, APL Mater. 4 (2016).

[24] F. Ren, L. Ward, T. Williams, K.J. Laws, C. Wolverton, J. Hattrick-Simpers, A. Mehta, Accelerated discovery of metallic glasses through iteration of machine learning and high-throughput experiments, Sci. Adv. 4 (2018) eaaq1566.

[25] P. Nikolaev, D. Hooper, F. Webber, R. Rao, K. Decker, M. Krein, J. Poleski, R. Barto, B. Maruyama, Autonomy in materials research: A case study in carbon nanotube growth, Npj Comput. Mater. 2 (2016).

[26] E. Kim, K. Huang, A. Saunders, A. McCallum, G. Ceder, E. Olivetti, Materials Synthesis Insights from Scientific Literature via Text Extraction and Machine Learning, Chem. Mater. 29 (2017) 9436–9444.

[27] S.R. Young, A. Maksov, M. Ziatdinov, Y. Cao, M. Burch, J. Balachandran, L. Li, S. Somnath, R.M. Patton, S. V. Kalinin, R.K. Vasudevan, Data mining for better material synthesis: The case of pulsed laser deposition of complex oxides, J. Appl. Phys. 123 (2018) 115303.

[28] J.S. Smith, B. Nebgen, N. Lubbers, O. Isayev, A.E. Roitberg, Less is more: Sampling chemical space with active learning, J. Chem. Phys. 148 (2018). doi:10.1063/1.5023802.

[29] T. Lookman, P. V. Balachandran, D. Xue, R. Yuan, Active learning in materials science with emphasis on adaptive sampling using uncertainties for targeted design, Npj Comput. Mater. 5 (2019) 21.

[30] A. Mannodi-Kanakkithodi, G. Pilania, T.D. Huan, T. Lookman, R. Ramprasad, Machine Learning Strategy for Accelerated Design of Polymer Dielectrics, Sci. Rep. 6 (2016).

[31] C. Kim, G. Pilania, R. Ramprasad, Machine Learning Assisted Predictions of Intrinsic Dielectric Breakdown Strength of ABX3 Perovskites, J. Phys. Chem. C. 120 (2016) 14575–14580.

[32] W. Li, R. Jacobs, D. Morgan, Predicting the thermodynamic stability of perovskite oxides using machine learning models, Comput. Mater. Sci. 150 (2018).

[33] G. Pilania, J.E. Gubernatis, T. Lookman, Multi-fidelity machine learning models for

accurate bandgap predictions of solids, Comput. Mater. Sci. 129 (2017) 156–163.

[34] R. Ramprasad, A. Mannodi-Kanakkithodi, T. Lookman, G. Pilania, B.P. Uberuaga, J.E. Gubernatis, Machine learning bandgaps of double perovskites, Sci. Rep. 6 (2016) 1–10.

[35] J. Lee, A. Seko, K. Shitara, K. Nakayama, I. Tanaka, Prediction model of band gap for inorganic compounds by combination of density functional theory calculations and machine learning techniques, Phys. Rev. B. 93 (2016) 115104.

[36] Y. Zhuo, A.M. Tehrani, J. Brgoch, Predicting the Band Gaps of Inorganic Solids by Machine Learning, J. Phys. Chem. Lett. 9 (2018) 1668–1673.

[37] V. Stanev, C. Oses, A.G. Kusne, E. Rodriguez, I. Takeuchi, S. Curtarolo, J. Paglione, Machine learning modeling of superconducting critical temperature, Npj Comput. Mater. 4 (2018).

[38] B. Meredig, E. Antono, C. Church, M. Hutchinson, J. Ling, S. Paradiso, B. Blaiszik, I. Foster, B. Gibbons, J. Hattrick-Simpers, A. Mehta, L. Ward, Can machine learning identify the next high-temperature superconductor? Examining extrapolation performance for materials discovery, Mol. Syst. Des. Eng. 3 (2018) 819–825.

[39] Y. -c. Liu, B. Afflerbach, R. Jacobs, S. -k. Lin, D. Morgan, Exploring effective charge in electromigration using machine learning, MRS Commun. 9 (2019) 567–575.

[40] H. Wu, A. Lorenson, B. Anderson, L. Witteman, H. Wu, B. Meredig, D. Morgan, Robust FCC solute diffusion predictions from ab-initio machine learning methods, Comput. Mater. Sci. 134 (2017) 160–165.

[41] H.-J. Lu, N. Zou, R. Jacobs, B. Afflerbach, X.-G. Lu, D. Morgan, Error assessment and optimal cross-validation approaches in machine learning applied to impurity diffusion, Comput. Mater. Sci. (2019) 109075.

[42] G. Pilania, K.J. McClellan, C.R. Stanek, B.P. Uberuaga, Physics-informed machine learning for inorganic scintillator discovery, J. Chem. Phys. 148 (2018).

[43] A. Seko, T. Maekawa, K. Tsuda, I. Tanaka, Machine learning with systematic density-functional theory calculations: Application to melting temperatures of single- and binary-component solids, Phys. Rev. B. 89 (2014).

[44] F. Pedregosa, G. Varoquaux, Scikit-learn: Machine learning in Python, 2011. doi:10.1007/s13398-014-0173-7.2.

[45] F. Chollet, Keras, (2015). https://github.com/keras-team/keras.

[46] M. Abadi, P. Barham, J. Chen, Z. Chen, A. Davis, J. Dean, M. Devin, S. Ghemawat, G. Irving, M. Isard, M. Kudlur, J. Levenberg, R. Monga, S. Moore, D.G. Murray, B. Steiner, P. Tucker, V. Vasudevan, P. Warden, M. Wicke, Y. Yu, X. Zheng, G. Brain, TensorFlow: A System for Large-Scale Machine Learning, in: 12th USENIX Symp. Oper. Syst. Des. Implement. (OSDI '16), 2016: pp. 265–284.

[47] University of Wisconsin-Madison MAST-ML development team, The MAterials Simulation Toolkit for Machine Learning (MAST-ML), (2018). https://github.com/uw-cmg/MAST-ML.

[48] M. Hutchinson, Citrine Informatics Lolo, 2016. https://github.com/CitrineInformatics/lolo.

[49] A. Tropsha, C. Toher, J. Carrete, E. Gossett, C. Oses, N. Mingo, F. Rose, F. Legrain, S. Curtarolo, E. Zurek, O. Isayev, AFLOW-ML: A RESTful API for machine-learning predictions of materials properties, Comput. Mater. Sci. 152 (2018) 134–145.

[50] L. Ward, A. Dunn, A. Faghaninia, N.E.R. Zimmermann, S. Bajaj, Q. Wang, J. Montoya, J. Chen, K. Bystrom, M. Dylla, K. Chard, M. Asta, K.A. Persson, G. Je, I. Foster, A. Jain, Matminer : An open source toolkit for materials data mining, Comput. Mater. Sci. 152 (2018) 60–69.

[51] D.B. Brough, D. Wheeler, S.R. Kalidindi, Materials Knowledge Systems in Python—a Data Science Framework for Accelerated Development of Hierarchical Materials, Integr. Mater. Manuf. Innov. 6 (2017) 36–53.

[52] veidt, (2015). https://github.com/materialsvirtuallab/veidt.

[53] L. Ward, C. Wolverton, Atomistic calculations and materials informatics : A review, Curr. Opin. Solid State Mater. Sci. 21 (2017) 167–176.

[54] L. Ward, A. Agrawal, A. Choudhary, C. Wolverton, A general-purpose machine learning framework for predicting properties of inorganic materials, Npj Comput. Mater. (2016) 1–7.

[55] S.P. Ong, W. Davidson, A. Jain, G. Hautier, M. Kocher, S. Cholia, D. Gunter, V.L. Chevrier, K.A. Persson, G. Ceder, Python Materials Genomics ( pymatgen ): A robust , open-source python library for materials analysis, Comput. Mater. Sci. 68 (2013) 314–319. doi:10.1016/j.commatsci.2012.10.028.

[56] A.P. Barok, R. Kondor, G. Csanyi, On representing chemical environments, Phys. Rev. B. 87 (2013) 184115.

[57] M. Rupp, A. Tkatchenko, K. Muller, O.A. Von Lilienfeld, Fast and Accurate Modeling of Molecular Atomization Energies with Machine Learning, Phys. Rev. Lett. 108 (2012) 58301.

[58] K. Hansen, F. Biegler, R. Ramakrishnan, W. Pronobis, O.A. von Lilienfeld, K.-R. Muller, A. Tkatchenko, Machine Learning Predictions of Molecular Properties : Accurate Many-Body Potentials and Nonlocality in Chemical Space, J. Phys. Chem. Lett. 6 (2015) 2326–2331.

[59] S. Raschka, mlxtend, (2019). https://github.com/rasbt/mlxtend.

[60] J. Ling, M. Hutchinson, E. Antono, S. Paradiso, B. Meredig, High-Dimensional Materials and Process Optimization Using Data-Driven Experimental Design with Well-Calibrated Uncertainty Estimates, Integr. Mater. Manuf. Innov. 6 (2017) 207–217. doi:10.1007/s40192-017-0098-z.

[61] S. Wager, T. Hastie, B. Efron, Confidence Intervals for Random Forests: The Jackknife and the Infintesimal Jackknife, J. Mach. Learn. Res. 15 (2014).

[62] M. Foord, N. Larosa, R. Dennis, E. Courtwright, configobj, 2014. https://github.com/DiffSK/configobj.